\def\ref#1{\lbrack #1\rbrack}

\def\eckk#1{\bigl[ #1 \bigr]}
\def\rund#1{\left( #1 \right)}
\def\abs#1{\left\vert #1 \right\vert}

\def\ave#1{\left\langle #1 \right\rangle}

\def\part#1#2{{\partial #1\over\partial #2}}

\def\d{{\rm d}}

\def\eps{{\epsilon}}

\def\vp{\varphi}

\def\ie{{\it i.e.~}}
\overfullrule=0pt
\def\eg{{\it e.g.~}}

\def\Real{{\rm I\mathchoice{\kern-0.70mm}{\kern-0.70mm}{\kern-0.65mm}%
  {\kern-0.50mm}R}}
\def\C{\rm C\kern-.42em\vrule width.03em height.58em depth-.02em
       \kern.4em}
\font \bolditalics = cmmib10
\def\bx#1{\leavevmode\thinspace\hbox{\vrule\vtop{\vbox{\hrule\kern1pt
        \hbox{\vphantom{\tt/}\thinspace{\bf#1}\thinspace}}
      \kern1pt\hrule}\vrule}\thinspace}

\def \vc #1{{\textfont1=\bolditalics \hbox{$\bf#1$}}}
{\catcode`\@=11
\gdef\SchlangeUnter#1#2{\lower2pt\vbox{\baselineskip 0pt \lineskip0pt
  \ialign{$\m@th#1\hfil##\hfil$\crcr#2\crcr\sim\crcr}}}
}
\def\gtrsim{\mathrel{\mathpalette\SchlangeUnter>}}
\def\lesssim{\mathrel{\mathpalette\SchlangeUnter<}}

 \def\etal   {{\it et~al.\ts}}
\def\ueber#1#2{{\setbox0=\hbox{$#1$}%
  \setbox1=\hbox to\wd0{\hss$\scriptscriptstyle #2$\hss}%
  \offinterlineskip
  \vbox{\box1\kern0.4mm\box0}}{}}

\def\bx#1{\leavevmode\thinspace\hbox{\vrule\vtop{\vbox{\hrule\kern1pt
        \hbox{\vphantom{\tt/}\thinspace{\bf#1}\thinspace}}
      \kern1pt\hrule}\vrule}\thinspace}

\def\SFB{{This work was supported by the ``Sonderforschungsbereich
375-95 f\"ur
Astro--Teil\-chen\-phy\-sik" der Deutschen For\-schungs\-ge\-mein\-schaft.}}
\magnification=\magstep1
\input epsf
\voffset= 0.0 true cm
\vsize=19.8 cm     
\hsize=13.5 cm
\hfuzz=2pt
\tolerance=500
\abovedisplayskip=3 mm plus6pt minus 4pt
\belowdisplayskip=3 mm plus6pt minus 4pt
\abovedisplayshortskip=0mm plus6pt
\belowdisplayshortskip=2 mm plus4pt minus 4pt
\predisplaypenalty=0
\footline={\tenrm\ifodd\pageno\hfil\folio\else\folio\hfil\fi}

\def\la{\mathrel{\hbox{\rlap{\hbox{\lower4pt\hbox{$\sim$}}}\hbox{$<$}}}}
\def\ga{\mathrel{\hbox{\rlap{\hbox{\lower4pt\hbox{$\sim$}}}\hbox{$>$}}}}

\def\arcmin{\hbox{$^\prime$}}
\def\arcsec{\hbox{$^{\prime\prime}$}}
\def\utw{\smash{\rlap{\lower5pt\hbox{$\sim$}}}}
\def\udtw{\smash{\rlap{\lower6pt\hbox{$\approx$}}}}

\def\getsto{\mathrel{\hbox{\rlap{$\gets$}\hbox{\raise2pt\hbox{$\to$}}}}}
\def\lid{\mathrel{\hbox{\rlap{\hbox{\lower4pt\hbox{$=$}}}\hbox{$<$}}}}
\def\gid{\mathrel{\hbox{\rlap{\hbox{\lower4pt\hbox{$=$}}}\hbox{$>$}}}}
\def\sol{\mathrel{\hbox{\rlap{\hbox{\raise4pt\hbox{$\sim$}}}\hbox{$<$}}}
}
\def\sog{\mathrel{\hbox{\rlap{\hbox{\raise4pt\hbox{$\sim$}}}\hbox{$>$}}}
}
\def\lse{\mathrel{\hbox{\rlap{\hbox{\raise4pt\hbox{$<$}}}\hbox{$\simeq$}
}}}
\def\gse{\mathrel{\hbox{\rlap{\hbox{\raise4pt\hbox{$>$}}}\hbox{$\simeq$}
}}}
\def\grole{\mathrel{\hbox{\lower2pt\hbox{$<$}}\kern-8pt
\hbox{\raise2pt\hbox{$>$}}}}
\def\leogr{\mathrel{\hbox{\lower2pt\hbox{$>$}}\kern-8pt
\hbox{\raise2pt\hbox{$<$}}}}
\def\loa{\mathrel{\hbox{\rlap{\hbox{\lower4pt\hbox{$\approx$}}}\hbox{$<$
}}}}
\def\goa{\mathrel{\hbox{\rlap{\hbox{\lower4pt\hbox{$\approx$}}}\hbox{$>$
}}}}

%
%

\font\kleinhalbcurs=cmmib10 scaled 833
\font\eightrm=cmr8
\font\sixrm=cmr6
\font\eighti=cmmi8
\font\sixi=cmmi6
\skewchar\eighti='177 \skewchar\sixi='177
\font\eightsy=cmsy8
\font\sixsy=cmsy6
\skewchar\eightsy='60 \skewchar\sixsy='60
\font\eightbf=cmbx8
\font\sixbf=cmbx6
\font\eighttt=cmtt8
\hyphenchar\eighttt=-1
\font\eightsl=cmsl8
\font\eightit=cmti8

\font\bxf=cmbx10
  \mathchardef\Gamma="0100
  \mathchardef\Delta="0101
  \mathchardef\Theta="0102
  \mathchardef\Lambda="0103
  \mathchardef\Xi="0104
  \mathchardef\Pi="0105
  \mathchardef\Sigma="0106
  \mathchardef\Upsilon="0107
  \mathchardef\Phi="0108
  \mathchardef\Psi="0109
  \mathchardef\Omega="010A
\def\rahmen#1{\vskip#1truecm}
\def\begfig#1cm#2\endfig{\par
\setbox1=\vbox{\rahmen{#1}#2}%
\dimen0=\ht1\advance\dimen0by\dp1\advance\dimen0by5\baselineskip
\advance\dimen0by0.4true cm
\ifdim\dimen0>\vsize\pageinsert\box1\vfill\endinsert
\else
\dimen0=\pagetotal\ifdim\dimen0<\pagegoal
\advance\dimen0by\ht1\advance\dimen0by\dp1\advance\dimen0by1.4true cm
\ifdim\dimen0>\vsize
\topinsert\box1\endinsert
\else\vskip1true cm\box1\vskip4true mm\fi
\else\vskip1true cm\box1\vskip4true mm\fi\fi}
\def\figure#1#2{\smallskip\setbox0=\vbox{\noindent\petit{\bf Fig.\ts#1.\
}\ignorespaces #2\smallskip
\count255=0\global\advance\count255by\prevgraf}%
\ifnum\count255>1\box0\else
\centerline{\petit{\bf Fig.\ts#1.\ }\ignorespaces#2}\smallskip\fi}

\def\xfigure#1#2#3#4{\midinsert\noindent
    $$\epsfxsize=#4truecm\epsffile{#3}$$
    \figure{#1}{#2}\endinsert}


\def\begtab#1cm#2\endtab{\par
\ifvoid\topins\midinsert\vbox{#2\rahmen{#1}}\endinsert
\else\topinsert\vbox{#2\kern#1true cm}\endinsert\fi}
\def\rahmen#1{\vskip#1truecm}
\def\begpet{\vskip6pt\bgroup\petit}
\def\endpet{\vskip6pt\egroup}
\def\begref{\par\bgroup\petit
\let\it=\rm\let\bf=\rm\let\sl=\rm\let\INS=N}
\def\petit{\def\rm{\fam0\eightrm}%
\textfont0=\eightrm \scriptfont0=\sixrm \scriptscriptfont0=\fiverm
 \textfont1=\eighti \scriptfont1=\sixi \scriptscriptfont1=\fivei
 \textfont2=\eightsy \scriptfont2=\sixsy \scriptscriptfont2=\fivesy
 \def\it{\fam\itfam\eightit}%
 \textfont\itfam=\eightit
 \def\sl{\fam\slfam\eightsl}%
 \textfont\slfam=\eightsl
 \def\bf{\fam\bffam\eightbf}%
 \textfont\bffam=\eightbf \scriptfont\bffam=\sixbf
 \scriptscriptfont\bffam=\fivebf
 \def\tt{\fam\ttfam\eighttt}%
 \textfont\ttfam=\eighttt
 \normalbaselineskip=9pt
 \setbox\strutbox=\hbox{\vrule height7pt depth2pt width0pt}%
 \normalbaselines\rm
\def\vec##1{\setbox0=\hbox{$##1$}\hbox{\hbox
to0pt{\copy0\hss}\kern0.45pt\box0}}}%
\let\ts=\thinspace
%
\font \tafontt=     cmbx10 scaled\magstep2
\font \tafonts=     cmbx7  scaled\magstep2
\font \tafontss=     cmbx5  scaled\magstep2
\font \tamt= cmmib10 scaled\magstep2
\font \tams= cmmib10 scaled\magstep1
\font \tamss= cmmib10
\font \tast= cmsy10 scaled\magstep2
\font \tass= cmsy7  scaled\magstep2
\font \tasss= cmsy5  scaled\magstep2
\font \tasyt= cmex10 scaled\magstep2
\font \tasys= cmex10 scaled\magstep1
\font \tbfontt=     cmbx10 scaled\magstep1
\font \tbfonts=     cmbx7  scaled\magstep1
\font \tbfontss=     cmbx5  scaled\magstep1
\font \tbst= cmsy10 scaled\magstep1
\font \tbss= cmsy7  scaled\magstep1
\font \tbsss= cmsy5  scaled\magstep1

\newbox\chsta\newbox\chstb\newbox\chstc
\def\centerpar#1{{\advance\hsize by-2\parindent
\rightskip=0pt plus 4em
\leftskip=0pt plus 4em
\parindent=0pt\setbox\chsta=\vbox{#1}%
\global\setbox\chstb=\vbox{\unvbox\chsta
\setbox\chstc=\lastbox
\line{\hfill\unhbox\chstc\unskip\unskip\unpenalty\hfill}}}%
\leftline{\kern\parindent\box\chstb}}
 \def \chap#1{
    \vskip24pt plus 6pt minus 4pt
    \bgroup
 \textfont0=\tafontt \scriptfont0=\tafonts \scriptscriptfont0=\tafontss
 \textfont1=\tamt \scriptfont1=\tams \scriptscriptfont1=\tamss
 \textfont2=\tast \scriptfont2=\tass \scriptscriptfont2=\tasss
 \textfont3=\tasyt \scriptfont3=\tasys \scriptscriptfont3=\tenex
     \baselineskip=18pt
     \lineskip=18pt
     \raggedright
     \pretolerance=10000
     \noindent
     \tafontt
     \ignorespaces#1\vskip7true mm plus6pt minus 4pt
     \egroup\noindent\ignorespaces}%
 \def \sec#1{
     \vskip25true pt plus4pt minus4pt
     \bgroup
 \textfont0=\tbfontt \scriptfont0=\tbfonts \scriptscriptfont0=\tbfontss
 \textfont1=\tams \scriptfont1=\tamss \scriptscriptfont1=\kleinhalbcurs
 \textfont2=\tbst \scriptfont2=\tbss \scriptscriptfont2=\tbsss
 \textfont3=\tasys \scriptfont3=\tenex \scriptscriptfont3=\tenex
     \baselineskip=16pt
     \lineskip=16pt
     \raggedright
     \pretolerance=10000
     \noindent
     \tbfontt
     \ignorespaces #1
     \vskip12true pt plus4pt minus4pt\egroup\noindent\ignorespaces}%
 \def \subs#1{
     \vskip15true pt plus 4pt minus4pt
     \bgroup
     \bxf
     \noindent
     \raggedright
     \pretolerance=10000
     \ignorespaces #1
     \vskip6true pt plus4pt minus4pt\egroup
     \noindent\ignorespaces}%
 \def \subsubs#1{
     \vskip15true pt plus 4pt minus 4pt
     \bgroup
     \bf
     \noindent
     \ignorespaces #1\unskip.\ \egroup
     \ignorespaces}
\def\footnoterule{\kern-3pt\hrule width 2true cm\kern2.6pt}
\newcount\footcount \footcount=0
\def\advftncnt{\advance\footcount by1\global\footcount=\footcount}
\def\fonote#1{\advftncnt$^{\the\footcount}$\begingroup\petit
       \def\textindent##1{\hang\noindent\hbox
       to\parindent{##1\hss}\ignorespaces}%
\vfootnote{$^{\the\footcount}$}{#1}\endgroup}

\newcount\sterne
\outer\def\byebye{\bigskip\typeset
\sterne=1\ifx\speciali\undefined\else
\bigskip Special caracters created by the author
\loop\smallskip\noindent special character No\number\sterne:
\csname special\romannumeral\sterne\endcsname
\advance\sterne by 1\global\sterne=\sterne
\ifnum\sterne<11\repeat\fi
\vfill\supereject\end}
\def\typeset{\centerline{\petit This article was processed by the author
using the \TeX\ Macropackage from Springer-Verlag.}}

\voffset=0pt

\def\L{{\cal L}}
\chap{Quantitative analysis of galaxy-galaxy lensing}
\centerline{Peter Schneider$^1$ and Hans-Walter Rix$^{1,2}$}
\medskip
\centerline{$^1$ Max-Planck-Institut f\"ur Astrophysik}
\centerline{Postfach 1523, D-85740 Garching, Germany}
\medskip
\centerline{$^2$ Steward Observatory}
\centerline{University of Arizona, Tucson, AZ 85721, USA}
\bigskip
\sec{Abstract}
Gravitational light deflection due to mass along the line-of-sight 
will distort the images of background sources. This effect
has been used successfully to investigate the mass distribution of
galaxy clusters. 
Although an individual galaxy is not massive
enough to cause a detectable lensing distortion in the background population,
this effect can be measured statistically 
for a population of galaxies,  and a first detection 
was reported recently by Brainerd, Blandford and Smail (BBS).

In this paper we explore a quantitative and efficient method to constrain the
halo properties of distant galaxy populations through
``galaxy--galaxy" lensing and show that the mean
masses and sizes of halos can be estimated accurately, without excessive
data requirements.  Specifically, we propose a
maximum-likelihood analysis which takes full account of the actual
image ellipticities, positions and apparent magnitudes.
We apply it to simulated observations, using
the same model for the lensing galaxy population
as in BBS, where the galaxy halos are described by
isothermal spheres with velocity dispersion
$\sigma$, truncated at a radius $s$. Both parameters are assumed to
scale with the luminosity of the galaxy. The best fitting values,
$\sigma_*$ and $s_*$, corresponding to an $L_*$-galaxy,
are then determined with the maximum-likelihood analysis.
We explore two different 
observing strategies, (a) taking deep images (e.g., with HST) on small
fields, and (b) using shallower images on larger fields. 

From these simulations we find
that $\sigma_*$ can be determined to $\lesssim$10\% accuracy if a sample of
about 5000 galaxies with measured ellipticities are available, down to
$R\lesssim 23$. The corresponding data can be obtained on a 4-meter
class telescope in a few nights of very good seeing. Alternatively,
the same accuracy in the determination of $\sigma_*$ can be achieved
from about ten, moderately deep WFPC2 fields, on which galaxy shapes can
be measured to about $R\sim 25$ and for which ground-based images
are available on which the WFPC2 fields are centered.
Firm lower limits can be set on the radial extent of the 
halo, but the maximal halo extent is poorly constrained.
We show that the likelihood approach can also be used to constrain
other parameters of the galaxy population, such as the Tully-Fischer
index, or the mean redshift of the galaxies as a function of apparent
magnitude. Finally we show how multi-color information, constraining
the redshift of individual galaxies, can dramatically improve the accuracy of the
parameter determination.
\vfill \eject

\sec{1 Introduction}
The distortion of distant source images by the gravitational
field of mass concentrations close to their line-of-sight
has long been recognized as a powerful tool for
studying the distribution of (dark) matter in the universe on various
scales (e.g., Kristian 1967; Blandford \& Jaroszy\'nski 1981; Webster
1985; Blandford et al.\ts 1991; Kaiser 1992). In particular, the
distortion of faint galaxy images by foreground clusters, either in
forming arcs or arclets (see the recent review by Fort \& Mellier
1994) or by using weaker distortions (Tyson, Valdes \& Wenk 1990;
Kaiser \& Squires 1993) has recently been used to investigate the
projected mass
distribution of selected clusters of galaxies (Bonnet, Mellier \& Fort
1994; Fahlman et al.\ts 1994;
Smail et al.\ts 1995; Seitz et al.\ts 1995; Squires et al.\ts 1995) or
to detect otherwise 
unseen mass concentrations (Bonnet et al.\ts 1993; Fort et al.\ts
1995). Distortions due to larger-scale mass distributions
have not yet been detected at a highly significant level (Mould et
al.\ts 1994), though a possible detection has been reported by
Villumsen (1995).

In this paper we explore how to best probe the mass
distribution in isolated foreground galaxies through
the lensing distortion of background galaxy images.
For the idealized case of an isolated, isothermal
halo (with critical radius $\theta_{\rm cr}$) this distortion is
merely a small stretching (by $\theta_{\rm cr}/\theta$)
of the background image at separation $\theta$, perpendicular to the
line connecting the source--lens pair.
The distortion caused by a single galaxy cannot be
detected, unless its velocity dispersion were greater than about
$\sigma\sim 400$\ts km/s (Mirald-Escud\'e 1991; Schneider \& Seitz
1995), but the statistics of many foreground~--~background
pairs should yield a detectable signal. 
Tyson et al.\ts (1984) were the first to search for
this effect, but failed to detect it. They translated their result
into an upper bound for the mass and extent of halos around
$L_*$-galaxies, which was revised upward significantly 
by Kovner \& Milgrom (1987) in a more realistic analysis of the
observational results.
Tyson et al.'s\ non-detection, despite the
very large number of galaxy images, apparently discouraged other
attempts for a decade.
However, Brainerd, Blandford \& Smail (1995; henceforth BBS)
have now discovered a significant galaxy-galaxy lensing signal (at
the 99\% confidence level) in the data of Mould et al.\ts (1994). 
The effect was found by considering the statistical distribution
of the angle between the line connecting foreground-background pairs
and the major axis of the background galaxy.
BBS considered pair separations between and $5\arcsec$ and $34\arcsec$
and split their sample at $m_r=23$, calling brighter galaxies 
`foreground' and fainter ones `background' objects.
They then performed detailed Monte-Carlo
simulations resembling their data, using a parameterized model for the
galaxy population (see Sect.\ts 2 below),
to verify that the observed alignment statistics
is compatible with expectations from their galaxy model.

The two most important parameters describing a galaxy halo are its
mass (or velocity dispersion, $\sigma$) and its radial extent.
In the local universe,
rotation curves of spiral galaxies suggest the
presence of a dark halo with roughly isothermal profile, but little is
known about the mass density of galaxies beyond $\sim 30$\ts kpc (see
Zaritsky \& White 1994 for the exception).
Galaxy-galaxy lensing can in principle probe
the size of galactic halos. However, the angular extent of a halo of
size $\sim 100$\ts kpc is considerably larger than the mean angular
separation of two `foreground' galaxies; therefore, more than one
lensing galaxy will be important for the distortion of a background image
(see the description of the role of multiple
deflectors in BBS). Hence, in order to probe large halo sizes, a
statistical analysis must be employed which accounts for collective
effects.  Finally, though the magnitude of a
galaxy is correlated with its redshift, ordering of galaxy
distances by their magnitudes is only a rough approximation. By
fitting alignment curves through their observational data, BBS
accounted for all these effects properly. However, they did not
attempt to fit model parameters of their galaxy population to their
data.

Here we suggest a new and efficient
statistical method for analyzing data on
galaxy-galaxy lensing. Its decisive advantage over previous
methods lies in the fact that it exploits
all the information of the {\it actual} image configuration
(each model predicting the shear for each individual galaxy image)
rather than using only the ensemble properties (\eg the mean
tangential alignement of major axes) of statistically equivalent samples.
It is a maximum-likelihood method which uses
the magnitudes, angular positions, and ellipticities of galaxy images
as input values. The redshift probability distribution of the
galaxies, $p_z(z|m)$, is assumed to be known as a function of apparent magnitude;
from redshift surveys, this can be measured for relatively bright galaxies
and can be reasonably extrapolated to slightly fainter magnitudes (in fact, we
will show that this redshift distribution can be
constrained from galaxy-galaxy lensing). One can then calculate for
each galaxy image the expectation value of the shear and its
dispersion. This shear, combined with the intrinsic ellipticity distribution of
the sources, predicts the probability distribution for the
observed ellipticities. These are used to calculate the likelihood
function, which is then maximized with respect to the parameters of
interest. This method is described in detail in Sect.\ts 2. We then
generate synthetic data samples, as described in Sect.\ts 3 to which
we apply the maximum-likelihood analysis. In most cases, we are
interested in the velocity dispersion 
$\sigma_*$ and the size $s_*$ of an $L_*$-galaxy.  With only 
moderately large data sets (which should become available very
soon), either on fairly small fields but going deep (e.g., a
collection of WFPC2 exposures), or wide-angle field images of moderate
depth, $\sigma_*$ can be determined very accurately, and lower limits
on $s_*$ can be derived. As argued above, it is much more difficult to
obtain upper bounds on $s_*$. We then investigate whether the of the
results for $\sigma_*$ and $s_*$ depend sensitively on the knowledge of
other galaxy properties, such as the Tully-Fischer index, or the mean
redshift as a function of apparent magnitude. We find that these
covariances are not very strong and that in fact these additional
parameters can be significantly constrained with the maximum-likelihood
method. We demonstrate that even an approximate knowledge of the
individual galaxy redshifts improve the accuracy of the
results considerably; sufficiently accurate redshitfs can be
obtained from multicolor photometry (Connolly et al.\ts 1995). We
discuss our results in Sect.\ts 5, in view of the fact that
large data sets useful for our method of analysis will soon become
available, due to the increasing number of useful WFPC2 exposures, and
the new generation of 10m-class telescopes and wide-field cameras.
\vfill\eject

\sec{2 Method of analysis}
In this section we describe the statistical methods for 
the analysis of galaxy-galaxy lensing. The basic strategy is outlined
in Sect.\ts 2.1, and more specific assumptions are given in
Sect.\ts2.2. 

\subs{2.1 The strategy}
We assume that the imaging data available are derived
from a single field, and consist of the locations $\vc\theta_i$,
and the magnitudes $m_i$, $1\le i\le N_g$, of $N_g$ galaxies brighter
than some threshold $m_{\rm lim}$.  Furthermore, we assume
that the ellipticities $\eps_i$ are measured for a subset
of the galaxies, e.g.,
for all galaxies brighter than some magnitude $m_{\rm shape}$. (We will
refer to this subset as `galaxies with shape information'.)
Throughout this paper, we use the ellipticity parameter
$\eps$, which is a complex number with phase $2\vp$, where $\vp$
describes its orientation, and with an absolute value which for an
elliptical source with axis ratio $r\le 1$ is
$\abs{\eps}=(1-r)/(1+r)$. In general, $\eps$ must be calculated from
the second brightness moments of a galaxy image; see, e.g., Schneider
(1995) and references therein.
The observed ellipticities reflect
the intrinsic surface brightness distribution of the sources, modified
slightly by the 
shear induced by gravitational deflection. Galaxy-galaxy lensing tries
to infer the latter contribution from the observations in order to put
constraints on the deflecting mass distribution. 

Obviously, we can only consider contributions to the light deflection
from the detected galaxies in the field;
later, we will comment on
contributions from fainter (unobserved) galaxies, and from larger
scale mass distributions like cosmological shear or clusters of
galaxies.
We also ignore possible effects of 
all galaxies beyond $\theta_{\rm max}$ from any given
galaxy image.
If we knew the redshifts of all the galaxies and had a
model for their (halo) mass distribution, we could predict the
lensing shear exactly for each galaxy with shape information
and thus reconstruct the intrinsic ellipticity. Requiring
that the intrinsic ellipticities are randomly oriented can then
constrain the parameters of the (lensing) mass model.
This basic strategy will now be outlined in more detail.

We consider only galaxies with shape information
that lie at least $\theta_{\rm max}$ away from the field boundary.
For those images we can predict
the shear
by summing up the shear contributions from all the galaxies within a
circle of angular radius $\theta_{\rm max}$:
$$
\gamma_i=\sum_{z_j<z_i} \gamma_{ij}\quad,
\eqno (2.1)
$$
where $\gamma_i$ is the shear at the $i$-th galaxy image, obtained by summing
the shear contributions $\gamma_{ij}$ from all galaxies with
redshift $z_j$ smaller than $z_i$. The straight sum in (2.1) is not an
exact expression for the resulting shear, as the shear contributions from
multiple gravitational deflections do not add linearly 
(e.g., Blandford \& Narayan 1986,
Schneider, Ehlers \& Falco 1992 -- hereafter SEF,
Chap.\ts 9). But in the case considered here all shear contributions
will be small and the approximation (2.1) applies. The shear,
$\gamma$, is taken as a complex number, using the same notation as in
Schneider (1995). Unfortunately, even for a given mass model of the
galaxies, the sum in (2.1) cannot be performed, because we do not know
the redshifts of all galaxies. Instead we assign a
redshift probability distribution $p_z(z|m)$ to galaxies of a given
magnitude $m$ and calculate the expectation value $\ave{\gamma_i}$ for the
shear of the $i$-th galaxy.
This average requires integrations over the projected (physical)
separations and redshifts and cannot be performed analytically, even
for the simplest mass distributions. Such
averaging is best performed by Monte-Carlo integration: for
each galaxy in the data field we draw a redshift $z_i$ from
the distribution $p_z(z|m_i)$ and calculate for each
galaxy with shape information the shear
$\gamma_i$ from (2.1). Repeating this process $N_{\rm MC}$ times, the
mean shear and its dispersion, 
$$
\ave{\gamma_i}={1\over N_{\rm MC}}
\sum_{\nu=1}^{N_{\rm MC}} \gamma_i^\nu
$$ 
and
$\sigma_{\gamma,i}^2= \ave{\rund{\gamma_i-\ave{\gamma_i}}^2}$ can be
calculated.

Let $p^{(\rm s)}(\eps^{(\rm s)})\,\d^2 \eps^{(\rm s)}$ be the
probability that the intrinsic (complex) ellipticity of a galaxy lies
within $\d^2 \eps^{(\rm s)}$ of $\eps^{(\rm s)}$. This ellipticity
distribution might depend on the redshift and brightness of the
galaxies, or on its color. For simplicity we assume here that
this distribution is the same for all galaxies. For
weak shear, the relation between intrinsic ellipticity and observed
ellipticity is approximately
$$
\eps=\eps^{(\rm s)}-\gamma\quad ,
\eqno (2.2)
$$
so that the probability density for the observed ellipticity reads
$$
p_\eps(\eps|\gamma)=p^{(\rm s)}(\eps + \gamma)\quad .
\eqno (2.3)
$$
>From the Monte-Carlo integration, which for each realization of the
galaxy redshifts yields a value $\gamma_i^\nu$, the effective
ellipticity probability distribution is a mean of (2.3) over all
realizations,
$$
\ave{p_i}(\eps_i)={1\over N_{\rm MC}}\sum_{\nu=1}^{N_{\rm MC}}
p^{(\rm s)}(\eps_i + \gamma_i^\nu)\quad .
\eqno (2.4)
$$
>From these probability densities, we can now construct the likelihood
function
$$
\L=\prod_i \ave{p_i}(\eps_i)\quad ,
\eqno (2.5)
$$
where the product extends over all galaxy images with shape information.
Maximizing this likelihood function with respect to the model
parameters yields their best fitting values.

\subs{2.2 Specific assumptions}
To carry out the program outlined in the preceding subsection, we have
to provide a parametrized description of the galaxy
population. In this parametrization, we
follow essentially the description given in BBS.

\subsubs{Number counts -- magnitude relation} We assume that the
number counts of galaxies follow a power-law distribution of the form
$$
{\d \log N(m)\over \d m} \propto 10^{\gamma m}\quad ,
\eqno (2.6)
$$
where the slope $\gamma$ depends on the waveband used and can be
determined empirically. Note that we need this distribution function
only for generating the synthetic data set, as described in Sect.\ts
3; it is not used in the maximum likelihood analysis.

\subsubs{Redshift distribution} Following BBS, we assume that the
redshift distribution of the galaxies is given as
$$
p_z(z|m)={\beta z^2 \exp\rund{-(z/z_0)^\beta} \over \Gamma(3/\beta)
z_0^3}\quad,
\eqno (2.7a)
$$
where $\Gamma(x)$ is the gamma function, and $z_0(m)$ is assumed to
depend linearly on the
apparent magnitude of the galaxies,
$$
z_0=k_z\eckk{z_m+z_m'(m-22)}\quad,
\eqno (2.7b)    
$$
where $k_z$, $z_m$ and $z_m'$ are constants.

\subsubs{Lens model} As in BBS, we describe the mass distribution of
the galaxies by the axially-symmetric surface mass density
$$
\Sigma(\xi)={\sigma^2\over 2 G \xi}\rund{1-{\xi\over
\sqrt{s^2+\xi^2}}}\quad,
\eqno (2.8)
$$
which corresponds to a singular isothermal sphere with characteristic
outer scale $s$. We will assume that $s$ scales
quadratically with $\sigma$,
$$
s=s_*\rund{\sigma\over\sigma_*}^2=:s\,\hat\sigma^2\quad,
\eqno (2.9)
$$
where $\sigma_*$ is the velocity dispersion of an $L_*$-galaxy. Using
the notation of SEF, we
obtain for the dimensionless surface mass 
density $\kappa$ of a galaxy at redshift $z_{\rm d}$ for a source at
redshift $z_{\rm s}$:
$$
\kappa(\theta)=4\pi \rund{\sigma_*\over c}^2\,\hat\sigma^2\,{r_{\rm
ds}\over r_{\rm s}}\,{1\over
2\theta}\rund{1-{\theta\over\sqrt{\theta^2 + \theta_s^2}}} \quad,
\eqno (2.10)
$$
where the $r$-factors are angular-diameter distances in units of
$c/H_0$, $\theta$ is the 
angular separation of the source image from the center of the
potential, and 
$$
\theta_s={s\over r_{\rm d} c/H_0}\quad .
$$
Defining 
$$
\theta_t={s_*\over (c/H_0)}\quad,
\eqno (2.11)
$$
one obtains 
$$
\theta_s={\hat\sigma^2\,\theta_t\over r_{\rm d}}\quad.
\eqno (2.12)
$$
From Eq.\ts(8.15) of SEF, the modulus of the shear is
$$
\abs{\gamma}=4\pi \rund{\sigma_*\over c}^2\,\hat\sigma^2\,{r_{\rm
ds}\over r_{\rm s}}\rund{{2\theta_t+\theta \over 2 \theta^2}
-{\theta^2+2\theta_t^2\over 2
\theta^2\sqrt{\theta^2+\theta_t^2}}}\quad ,
\eqno (2.13)
$$
and the phase of $\gamma$ is determined by the direction of the
background galaxy image relative to the deflector center. Throughout
this paper we assume an Einstein-de Sitter universe, for which
$$\eqalign{
r_{\rm d}&={2\over 1+z_{\rm d}}\rund{1-{1\over \sqrt{1+z_{\rm
d}}}}\quad, \cr
{r_{\rm ds}\over r_{\rm s}}&={\sqrt{1+z_{\rm s}} -\sqrt{1+z_{\rm d}}
\over \rund{\sqrt{1+z_{\rm s}} -1} \sqrt{1+z_{\rm d}}}\quad . \cr }
\eqno(2.14)
$$

\subsubs{Velocity dispersion -- luminosity relation} To link the
dynamical halo properties to the observable galaxy luminosity,
we use a Faber--Jackson~/~Tully--Fischer type relation of the form
$$
{L\over L_*}=\rund{\sigma\over \sigma_*}^\eta\quad .
\eqno (2.15)
$$

\subsubs{Luminosity-magnitude relation} Again following BBS, we shall
take
$$
{L\over L_*}=r_{\rm d}^2\,(1+z)^{3+\alpha}\,10^{0.4(23.9-m)}\quad,
\eqno (2.16)
$$
where the numerical value in the exponent was chosen to apply to
photometry in the red band, and $\alpha$ accounts for the
k-correction. 

\subsubs{Intrinsic ellipticity distribution} The intrinsic ellipticity
distribution of galaxies can be determined in principle from
high-resolution imaging; for example, the HST Medium Deep Survey will
most likely provide an accurate description of the intrinsic
ellipticity distribution. In this paper, we shall for simplicity
assume a distribution of the form
$$
p^{(\rm s)}(\eps^{(\rm s)})={1\over \pi \rho^2}
\exp\rund{-{\abs{\eps^{(\rm
s)}}^2\over \rho^2}}\quad .
\eqno (2.17)
$$
Then the averaged probability distribution for the observed
ellipticities becomes
$$
\ave{p_i}(\eps_i)={1\over \pi\,\rho^2\,N_{\rm MC}}\sum_{\nu=1}^{N_{\rm MC}}
\exp\rund{-{\abs{\eps_i+\gamma_i^\nu}^2\over \rho^2}}\quad .
\eqno (2.18)
$$
Since the shear will be small, the resulting probability distribution
will again approximately be a Gaussian,
$$
\ave{p_i}(\eps_i)={1\over \pi\,(\rho^2+\sigma_{\gamma,i}^2)}
\exp\rund{-{\abs{\eps_i+\ave{\gamma_i}}^2\over
\rho^2+\sigma_{\gamma,i}^2}}\quad ,
\eqno (2.19)
$$
in which case the log-likelihood function takes the form
$$
\ell:=\ln \L=-\sum_i {\abs{\eps_i+\ave{\gamma_i}}^2\over
\rho^2+\sigma_{\gamma,i}^2}-\sum_i
\ln\eckk{\pi\,(\rho^2+\sigma_{\gamma,i}^2)}
\quad .
\eqno (2.20)
$$
If $\ell_{\rm max}$ is the maximum value of the log-likelihood, then
the probability distribution of $2(\ell_{\rm max}-\ell)$ follows
approximately a $\chi^2$ distribution with the numbers of degrees
freedom equal to the number of varied model parameters.

\subsubs{Fiducial values of the parameters} Some of the parameters
necessary to specify the various distributions can be easily
determined from existing or forthcoming observations, and are thus not
critical. For example, the ellipticity distribution of the source images
will become known very soon. Unless noted otherwise, we shall use 
the following values of the parameters:
$\alpha=3$, $\eta=4$, $\beta=1.5$, $k_z=0.7$, $z_m=0.47$, $z_m'=0.1$,
$\rho=0.2$.

\sec{3 Simulations}
The subsequent simulations are done by distributing the galaxies
in a single quadratic field of sidelength $\theta_{\rm field}$.
Again, galaxies with shape information are only considered in the likelihood
function if they are at least $\theta_{\rm max}$ from the
the field boundary. It is clear that the method does not require a single contiguous
field; what matters is the number of galaxies with shape information for
which the location of
surrounding galaxies (within $\theta_{\rm max}$) are known. For
example, one could consider a number of WFPC2 fields for which
ground-based images are available on which the WFPC2 field is
centered; this would yield essentially equivalent information as the
one synthetically generated here.

The test of the maximum likelihood algorithm consists of two steps,
(1) the generation of ``data", drawn from an input model for the galaxy
population, and (2) the estimate of the input parameters by calculating
the relative likelihoods of different retrieving models.
In the simulations we generate data by distributing galaxies
over a quadratic field of size $\theta_{\rm field}+2\theta_{\rm max}'$,
with a magnitude distribution of the form (2.6), and with $\gamma=0.3$, such
that the surface number density of galaxies is 70/sq.arcmin
($m\le 26$, Smail \etal 1995). 
The galaxies are distributed in the magnitude interval $m\in[20,26]$,
and are assigned a redshift according to the distribution
(2.7).  We take $\theta_{\rm max}'=5$\ts arcmin.
We choose a set of input values for $\sigma_*$ and $s_*$,
$\sigma_*^{\rm in}$ and $s_*^{\rm in}$, 
calculate the shear from (2.1) for each galaxy within the
square of size $\theta_{\rm field}$ and 
draw an intrinsic ellipticity from the probability distribution 
(2.17). From (2.2) the observed ellipticity of each
galaxy is then calculated and stored along with the angular
positions and magnitudes of the galaxies.

To retrieve the input parameters $\sigma_*^{\rm in}$ and $s_*^{\rm in}$
for the lens model, we proceed as
described in the previous section. 
We assume that galaxies brighter than
$m_{\rm shape}$ carry shape information and that all
galaxies brighter than $m_{\rm lim}$ are observed. We perform the
likelihood analysis for all galaxies with shape information inside the
square of size $\theta_{\rm field}-2\theta_{\rm max}$ which have no
observed galaxy closer than $\theta_{\rm min}$, since for those the
ellipticity  will be difficult to obtain. For the Monte-Carlo
integration of $\ave{\gamma_i}$, typically $N_{\rm MC}=20$ is
sufficient. For all simulations presented in this paper, we choose 
$\sigma_*^{\rm in}=160$\ts km/s and $s_*^{\rm in}=80 h^{-1}$\ts kpc,
where $h$ is the Hubble constant in units
of 100\ts km/s/Mpc.
\vfill\eject

\sec{4 Results}

In the analysis at hand two sources of uncertainty must be considered:
First, purely statistical errors caused by the incompleteness of the
information available. In principle, these errors can always be decreased by using
larger and larger data sets. Second, even with unlimited data one may
arrive at the ``wrong answer", if the input model (or nature) is not included
in the space spanned by the models used in the likelihood analysis.
These two effects will be investigated in turn in Sections
\ts4.1 and \ts4.2 .
Finally, we shall demonstrate in Sect.\ts4.3 how much
better $\sigma_*^{\rm in}$ and $s_*^{\rm in}$ can be reconstructed,
if approximate redshifts ($\Delta z \sim 0.05$)for the galaxies are available.

\subs{4.1 Correct model assumptions}
Fig.\ts 1 shows the results of the likelihood analysis, for
different assumptions about the amount and the quality
of the available data. Contours of
constant log-likelihood $\ell$ as a function of
the trial parameters $s_*$ and $\sigma_*$ in the reconstruction are
displayed. 
The four panels represent results for two different 
threshold magnitudes $m_{\rm shape}$ and data field sizes
$\theta_{\rm field}$.
\xfigure{1}{Log-likelihood contours in the $s_*$-$\sigma_*$ plane, for
$\theta_{\rm field}=10'$ (two upper panels) and $\theta_{\rm
field}=15'$ (lower panels). In the left panels, $m_{\rm shape}=24$, in
the right panels $m_{\rm shape}=24.5$. The number of galaxies whose
shape information has been used, i.e., which are brighter than $m_{\rm
shape}$, have no `visible' galaxy inside $\theta_{\rm min}=3\arcsec$,
and are at least $\theta_{\rm max}=1\arcmin$ away from the boundary of
the field, is 795 (upper left panel), 1165 (upper right panel), 2169
(lower left panel), and 3137 (lower right panel). All galaxies
brighter than $m_{\rm lim}=25$ are taken into account for calculating
the mean shear $\ave{\gamma_i}$ and its dispersion $\sigma_{\gamma,i}$
according to (2.1).  The contours in this and the following
contour plots are $\ell_{\rm max}-\ell=0.1, 0.5, 1., 1.5
\dots$. The $O$ marks the input values of the parameters ($s_*^{\rm in}=80
h^{-1}$\ts kpc, $\sigma_*^{\rm in}=160$\ts km/s), and $X$ denotes the
position of the maximum of the likelihood function inside the displayed
parameter space}{fig1.tps}{13}
%
%
For all parameters the log-likelihood function has a broad
maximum (X), which is elongated in the direction of
$s_*$, because $s_*$ is much less well determined than
$\sigma_*$. Note that a difference of  $\ell_{\rm max}-\ell=1,2,3$
corresponds to a confidence level of $p=63\%, 86\%, 95\%$, 
respectively, implying that the retrieved parameter values at the likelihood 
maximum are statistically consistent with
the input value of $(s_*,\sigma_*)$.
The figure also shows that the upper bound on $s_*$ is unconstrained
in these particular simulations,
but that  significant lower bounds on $s_*$ can be
obtained for three out of four cases.
In addition, Figure 1 illustrates that the velocity dispersion $\sigma_*$ is well
constrained by this method. As expected, the accuracy with which the
model parameters can be derived increases dramatically with the number
of galaxy images used for the analysis.

Each panel in Figure 1 shows the results for one particular 
realization of the data set. To determine how well, on average,
the lens parameters $\sigma_*$ and $s_*$ can be determined, the analysis shown
in each panel should be repeated with other data realizations,
producing many likelihood maps. To present all this information
we resort to the following approach:
for any assumed value of $s_*$, one can determine the value of
$\sigma_*$ which maximizes the likelihood function, denoted by
$\hat\sigma_*(s_*)$. The set $\hat\sigma_*(s_*)$ traces a curve in the
$(s_*,\sigma_*)$-plane, which contains 
the best fit values of $(s_*,\sigma_*)$ for the given
data set. For the same parameters as those used in Fig.\ts 1, we have
calculated these curves for ten realizations of the galaxy
distribution; they are plotted in Fig.\ts2.
\xfigure{2}{The value $\hat\sigma_*(s_*)$ which maximizes the
likelihood for a fixed value of $s_*$ is plotted for ten different
realizations of the galaxy distribution. The parameters for each panel
are the same as the corresponding ones in Fig.\ts1. The filled hexagon
on each curve denotes the point where the likelihood attains its
maximum, within the range $10 h^{-1}{\rm kpc}\le s_*\le 150 h^{-1}{\rm
kpc}$. In the upper left panel, the lowest (dashed) curve attains its
minimum of the likelihood at $\sigma_*=124$\ts km/s.
The triangles, stars, and squares correspond to points on the
curves where $\ell_{\rm max}-\ell(s_*,\hat\sigma_*)=1,2,3$,
respectively} {fig2.tps}{13}
Although this figure appears crowded, it is worthwhile to discuss
it in some detail. As an example, consider the upper right panel. In
all ten cases, the maximum of the likelihood is attained at  values of
$\sigma_*$ (filled hexagons) which are less than $\sim 25$\ts km/s
away from the input 
value of 160\ts km/s. In fact, in eight of these ten cases, the
difference of the best fit value from the input value is less than
10\ts km/s. The value of $s_*$ is not well constrained, as was already
clear from the contour maps in Fig.\ts 1. On the other hand, in all
ten cases can a lower limit on $s_*$ be determined. The points on the
$\hat\sigma_*(s_*)$ curves where $\ell_{\rm max}-\ell=1,2,3$ are marked
with triangles, stars, and squares, respectively. 
In none of the ten
cases does the 68\% lower limit of $s_*$ lie above the true
value of $s_*$. In seven (six) cases, even 86\% (95\%) confidence
upper limits can be obtained. The figure furthermore shows that the
information which can be extracted from a synthetic data set varies
from realization to realization, but that the significance of the
values obtained can be extracted from the data by our statistical
approach. 

The decrease of $\hat\sigma_*(s_*)$ with $s_*$ shows that the
best-determined value from the likelihood analysis is a mass
estimate. The mass of an $L_*$-galaxy within a projected radius $\xi$
is
$$
M(<\xi)={\pi \sigma_*^2\over 2
G}\eckk{s_*+\xi-\sqrt{s_*^2+\xi^2}}\quad.
\eqno (4.1)
$$
One can now insert the value $\hat\sigma_*^2(s_*)$ into this mass
formula, which yields $M(<\xi)$ as a function of $s_*$. It turns out
that for the cases shown in Figs.\ts 1 and 2 this curve is fairly
flat for $\xi\sim 10h^{-1}$\ts kpc, and that the corresponding values
of $M(<\xi)$ lie within $\sim 15\%$ of the true value. Note that $10
h^{-1}$\ts kpc corresponds to about $\theta_{\rm min}=3\arcsec$ for a
galaxy at a typical redshift of $0.4$. 

In order to investigate the dependence of the results on the two
scales $\theta_{\rm min}$ and $\theta_{\rm max}$, we have plotted in
Fig.\ts 3 the likelihood contours for the same parameters as in the
upper right panel of Fig.\ts 1, except that we changed the values of
$\theta_{\rm min}$ and $\theta_{\rm max}$. 
\xfigure{3}{Same as Fig.\ts1, for the same parameters as those of the
upper right panel of Fig.\ts1, i.e., $\theta_{\rm field}=10'$ and
$m_{\rm shape}=24.5$, except that the inner and outer radii are
varied: upper left panel: 
$\theta_{\rm min}=3\arcsec$, $\theta_{\rm max} = 30\arcsec$, number of
galaxies with shape information used, $N=1471$;
upper right panel:
$\theta_{\rm min}=3\arcsec$, $\theta_{\rm max} = 120\arcsec$, $N=633$;
lower left panel:
$\theta_{\rm min}=2\arcsec$, $\theta_{\rm max} = 30\arcsec$, $N=1728$;
lower right panel:
$\theta_{\rm min}=2\arcsec$, $\theta_{\rm max} = 60\arcsec$, $N=1366$.
As before, the $O$ marks the input values of the parameters ($s_*^{\rm in}=80
h^{-1}$\ts kpc, $\sigma_*^{\rm in}=160$\ts km/s), and $X$ denotes the position
of the maximum of the likelihood function}{fig3.tps}{13}
For the same value of $\theta_{\rm min}$
and the same data field size, the difference between the two
right and the two left panels in Fig.\ts 3 lies in an increase
of $\theta_{\rm max}$. This has two effects: for each image with shape
information the number of potential lenses considered
increases and so does the information used in the likelihood analysis.
On the other hand, the increase in $\theta_{\rm max}$ reduces the area
where galaxy images with shape information are far enough from the field
edge to be considered in the analysis; this reduces $N$.
As can be seen from the figure,
the second effect dominates the first one, reiterating the importance
of the total galaxy image number on the accuracy of the results.
In addition, the reduction
of $\theta_{\rm min}$ in the lower two panels leads to a increase of
accuracy in the determination of $\sigma_*$.

\subsubs{Possible Applications}

In the following we will consider two specific types of data sets
to constrain the lens parameters $s_*$ and $\sigma_*$:
(1) HST images, where the
ellipticities can be measured accurately for fairly faint and 
small galaxies, i.e., where the number density of galaxies with shape
information is high. (2) Ground-based images, where the limit for shape
measurements is brighter, but where the reduced galaxy density
can be compensated by observations over a larger area on the sky.
\vfill\eject

\smallskip\centerline {\sl Deep Images / Small Fields}\smallskip

For the case (1) we consider, as before, a galaxy distribution with
70 objects per square arcminute ($20\le m\le 26$)
and the same input parameters as above.
We assume that the faint galaxies with shape information
come from a $7\arcmin\times 7\arcmin$ region, corresponding to
about 10 WFPC2 fields. In the following we choose
$\theta_{\rm field}=7\arcmin+2\theta_{\rm max}$, assuming that the
HST images will be supplemented by ground-based data, which cover a
larger area than the WFPC2 field.
Such data would provide shape and photometric information
for all galaxies in the WFPC2 field with
$m<m_{\rm shape}$; the ground-based data provide photometric
information for galaxies brighter than $m_{\rm limit}$,
which might be brighter than $m_{\rm shape}$. Those galaxies
brighter than $m_{\rm lim}$ within $\theta_{\rm max}$ are then used in
the likelihood analysis, as before.

\xfigure{4}{Log-likelihood contours, as in Fig.\ts1, for a particular
realization of the galaxy distribution,
for different combinations of the
parameter $\theta_{\rm 
min}$, $\theta_{\rm max}$, $m_{\rm shape}$ and $m_{\rm lim}$, as
indicated. Again, X denotes the maximum of the log-likelihood, O
denotes the input model values. In all cases, the solid angle from
which galaxies with shape information are taken is 49\ts
sq.arcmin., corresponding to about 10 WFC fields}{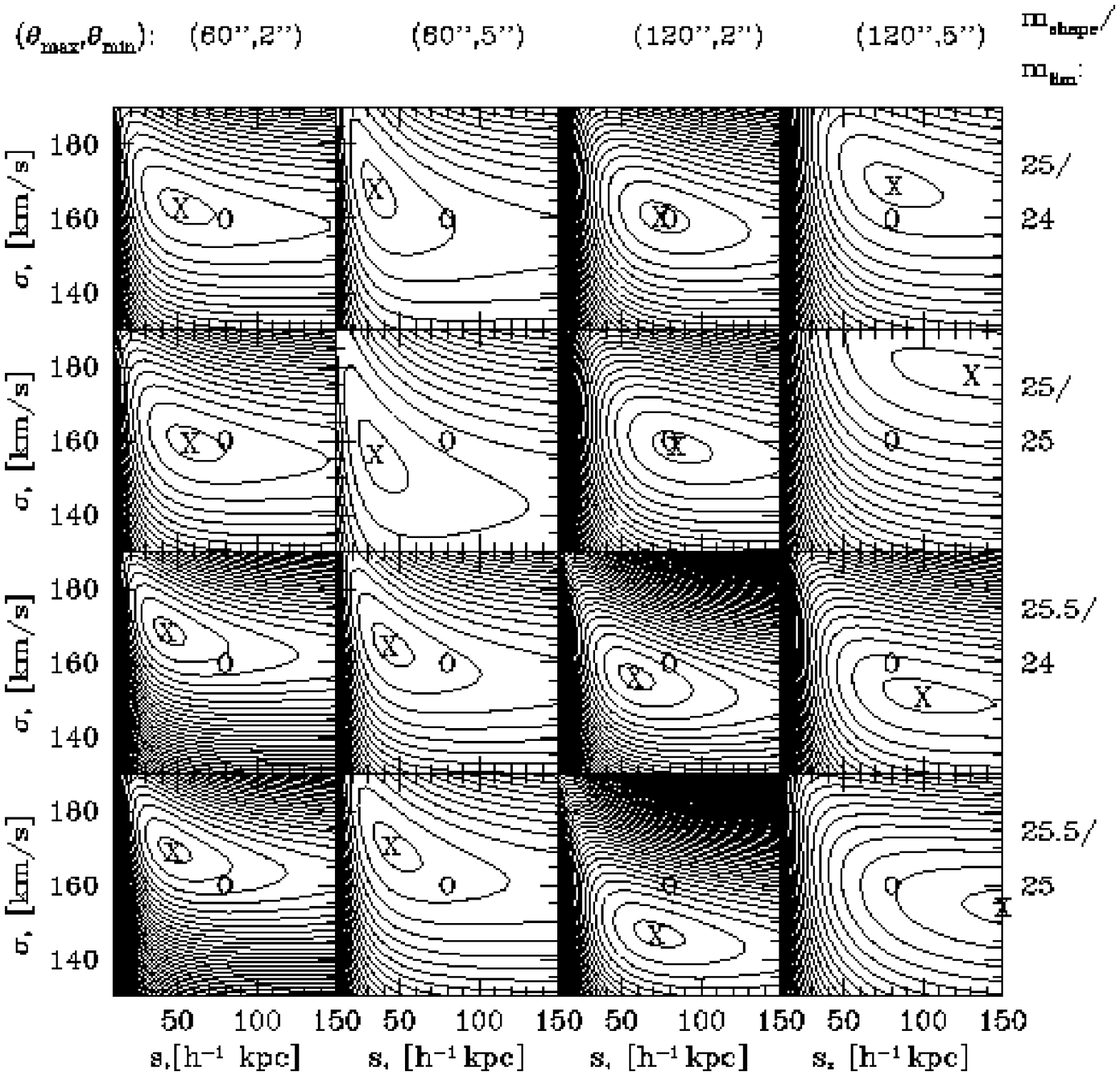}{18} 

\xfigure{5}{For different combinations of the parameters $\theta_{\rm
min}$, $\theta_{\rm max}$, $m_{\rm shape}$ and $m_{\rm lim}$, as
indicated, we have plotted for 30 realizations of the galaxy
distribution the best-fitting values for $s_*$ and $\sigma_*$, for the
same input parameters as before, i.e., $s_*^{\rm in}=80 h^{-1}$\ts kpc,
$\sigma_*^{\rm in}=160$\ts km/s. Values of $s_*>150 h^{-1}$\ts kpc do not
correspond to actual maxima of the log-likelihood function, but are
estimates of a lower limit of these maxima, obtained from
extrapolations of the likelihood function along curves
$\hat\sigma_*(s_*)$  such as plotted in Fig.\ts 2}{fig5.tps}{18}

In Fig.\ts4 we have plotted the log-likelihood contours in the
$(s_*,\sigma_*)$-plane, for several combinations of $\theta_{\rm
min}$, $\theta_{\rm max}$, $m_{\rm shape}$ and $m_{\rm lim}$. The
synthetic galaxy catalog is the same for all eight panels with the
same value of $\theta_{\rm max}$. Several general trends 
can be deduced: (1) decreasing $\theta_{\rm min}$ yields a
better defined likelihood maximum, because
tight source-lens pairs yield more information about $\sigma_*$ than
more distant pairs. (2) Considering fainter galaxies as potential
lenses (\ie increasing $m_{\rm lim}$)
does not improve the constraints on $s_*$ and $\sigma_*$,
since these fainter galaxies are mainly at redshifts
beyond the source galaxies'. (3) going to fainter
limits of $m_{\rm shape}$, however, provides tighter parameter
constraints by increasing the number of galaxies with shape information.
(4) an increase of the radius $\theta_{\rm max}$, within
which potential lenses are considered, improves the lower limits
on the halo cut-off $s_*$. This is intuitive,
because only the large-separation pairs can probe the cut-off
radius of the mass distribution of the lenses.
\xfigure{6}{Log-likelihood contours in the $s_*$-$\sigma_*$ plane, for
$\theta_{\rm field}=40'$. In the upper panels, $m_{\rm shape}=22.5$, in
the lower panels $m_{\rm shape}=23.5$. Only those galaxies without a
neighboring galaxy within $\theta_{\rm min}=3\arcsec$ (left panels)
and $\theta_{\rm min}=5\arcsec$ (right panels) were considered. In all
cases, $m_{\rm lim}=24$ and $\theta_{\rm max}=2\arcmin$.
The number of galaxies whose 
shape information has been used is
5891 (upper left panel), 4731 (upper right panel), 12975 (lower left
panel), and 10392 (lower right panel)}{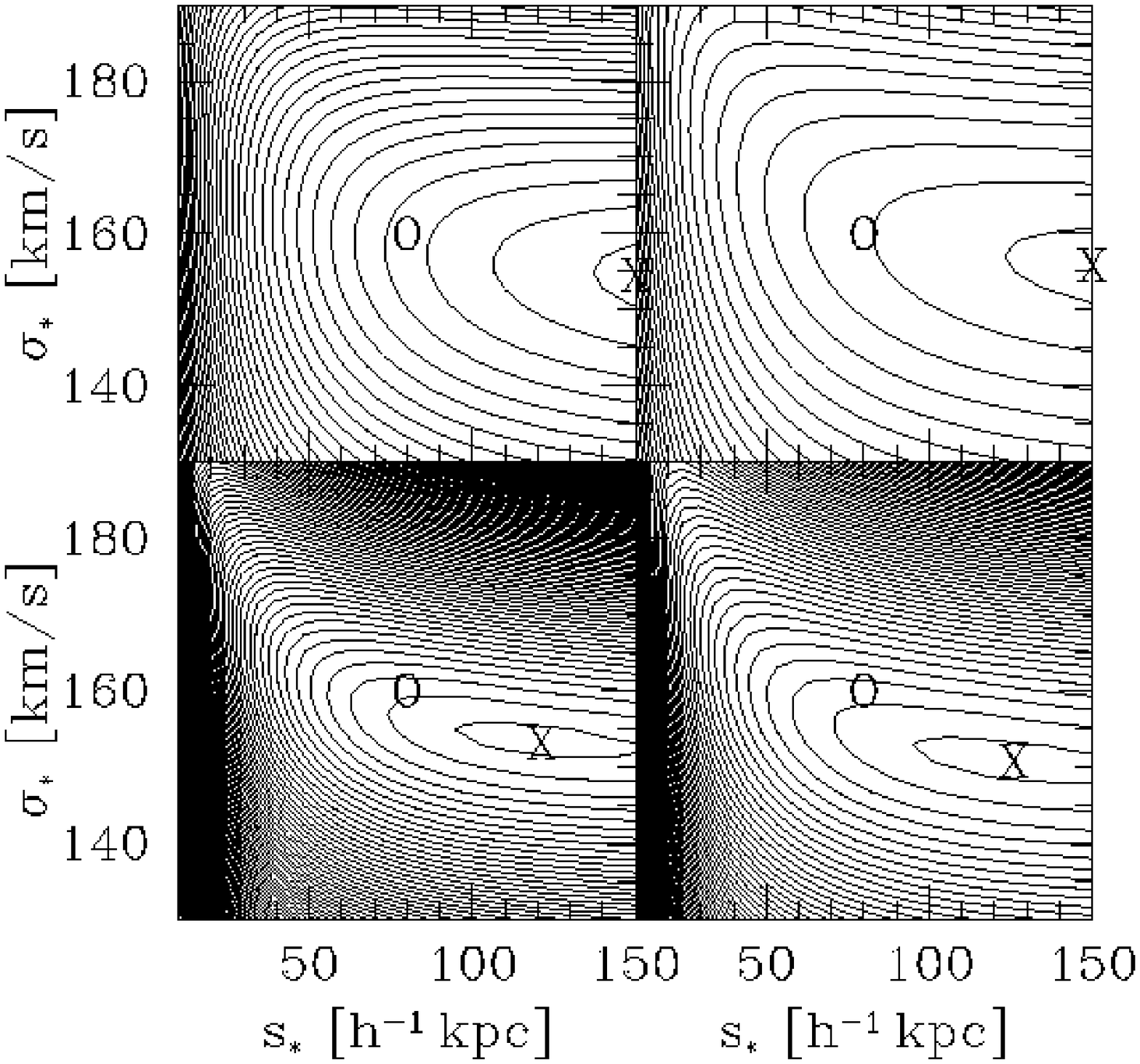}{13}
For the same combinations of parameters as in Fig.\ts 4, we have
repeated the calculation of the log-likelihood function for 30
realizations of the galaxy distribution, and determined the maximum of
$\ell$ as a function of $s_*$ and $\sigma_*$. The positions of these
maxima are plotted in Fig.\ts 5. As we have seen before, the maximum
of the likelihood function can occur at very large values of $s_*$,
since the upper limit on the cut-off radius is not well determined in many cases.
In order to display all points in a
limited region of the $(s_*,\sigma_*)$-plane, we have moved points
with $s_*>200 h^{-1}$\ts kpc somewhat arbitrarily to smaller values of
$s_*$; these still yield a fairly accurate determination of the
$\sigma_*$-value of the maximum.

Most importantly, the location of the log-likelihood maximum 
constitutes a fairly unbiased estimate of $\sigma_*$, except for
the rare cases where the estimate of $s_*$ is very small.
If the location of the
maximum of the log-likelihood function occurs at small values of
$s_*$, then these points tend to lie along curves similar to those
plotted in Fig.\ts 2, i.e., small values of $s_*$ correspond to large
values of $\sigma_*$. As we have discussed above, if the radius
$\theta_{\rm max}$ within which potential lenses are considered is
increased, small estimates of $s_*$ do no longer occur, i.e., much
better lower limits on $s_*$ can be obtained. Furthermore we see that
the scatter of points in Fig.\ts 5 is reduced when $\theta_{\rm min}$
is decreased and $m_{\rm shape}$ is increased.

\smallskip\centerline {\sl Less Deep Imaging / Wide Fields}\smallskip

We next consider the case of data on a large angular field,
but with a brighter limiting magnitude.
This case is more appropriate for ground-based
images, where the atmospheric effects limit the accuracy with which
ellipticities of faint and small galaxy images can be measured. A
galaxy distribution was simulated on a field of size $(40\arcmin)^2$
in the same way as before, i.e., 70 galaxies per sq.\ts arcmin.\ts brighter
than $m=26$ were
distributed and the ellipticities calculated. The log-likelihood
contours for four combinations of $\theta_{\rm min}$ and $m_{\rm
shape}$ are plotted in Fig.\ts 6; in all cases, only galaxies brighter
than $m_{\rm lim}=24$ were taken into account as potential lenses
within $\theta_{\rm max}=2\arcmin$. Whereas the galaxy sample is
considerably shallower than that in Fig.\ts 4, this effect is
compensated by the much larger solid angle: in all cases displayed, a
good approximation for $\sigma_*$ is obtained, as well as useful lower
bounds on $s_*$.

To summarize this subsection, we have shown that statistical
properties of the galaxy population can be inferred from galaxy-galaxy
lensing. Two strategies have been considered in somewhat more detail:
deep images of a fairly small field, or shallower images on large
fields. In both cases can significant constraints on the values of
$\sigma_*$ and $s_*$ be obtained. However, the information obtained in
both cases is not necessarily equivalent, since the deep images may reveal
effects of cosmological evolution of the galaxy population. The fact
that our analysis for the cases shown in Fig.\ts 6 yields estimates
for $\sigma_*$ and $s_*$ close to the input values shows that
additional galaxies which were not accounted for in our analysis
(namely those with $m_{\rm lim}=24\le m\le 26$) do not significantly
bias the estimate. The
analysis here has assumed that the redshift distribution of the
galaxies, the Tully-Fisher index, and the k-correction are known
precisely. In the next subsection, we shall investigate whether these
assumptions are critical, or whether some of these parameters can even
be constrained from galaxy-galaxy lensing.

\subs{4.2 Variation of the model parameters}
Next we consider the consequences of the fact that in practice
further parameters of the input model are unknown (not only
$\sigma_*$ and $s_*$). 
Instead of trying to cover a large region of parameter
space, we will focus on two parameters, the ``Tully-Fischer index"
$\eta$ and the parameter $z_m'$ (from Eq. 2.7),
which characterizes the change in the mean
redshift as a function of apparent magnitude.
\xfigure{7}{For the same data as used for Fig.\ts 6, the likelihood was
maximized as a function of $\sigma_*$, for three values of $s_*$:
$s_*=30$ 80, 130$h^{-1}$\ts kpc (solid, dotted, and dashed curves,
respectively). Here, $\theta_{\rm min}=3\arcsec$, $m_{\rm
shape}=23.5$, and $m_{\rm lim}=24$, i.e., these parameters correspond
to the lower left panel in Fig.\ts 6. The likelihood was maximized for
various values of the Tully-Fischer index $\eta$ (upper panel) and the
redshift parameter $z_m'$ occurring in eq.\ts(2.7) (lower panel). The
curves without crosses display the difference between the
log-likelihood of the best-fitting model and the log-likelihood of the
input model, i.e., for $\eta=4$, $z_m'=0.1$, $s_*=80 h^{-1}$\ts kpc,
and $\sigma_*=160$\ts km/s, whereas the curves with crosses display
the corresponding best-fitting value of $\sigma_*$}
{fig7.tps}{12}
We use the same synthetic data as used for
Fig.\ts 6, but calculate the likelihood function (2.20) for various
values of $\eta$ and $z_m'$. For three different values of
$s_*$ we specify a grid in $\eta$ and
$z_m'$ and maximize $\ell$ by varying $\sigma_*$ (\ie 
determine $\hat\sigma_*$ in the notation used in Fig.\ts 2).
In Fig.\ts 7 we plot the difference $\ell-\ell_{\rm true}$
between the log-likelihood of a model and its value
at the input parameters, as a function
of $\eta$ (upper panel) and as a function of $z_m'$ (lower
panel). Also, the corresponding value of $\sigma_*$ which maximizes
$\ell$ (for fixed $s_*$) is plotted. 

These simulations illustrate a number points. 
Whereas the log-likelihood curves are fairly flat
for values of $\eta\gtrsim4$, any value $\eta\lesssim 3$
can be excluded from the upper panel in Fig.\ts 7.
In the limit $\eta\rightarrow 0$ all the lensing power would be
concentrated in the very brightest galaxies; this can be rejected by
this analysis. The weak constraint on $\eta\rightarrow\infty$
does not imply a correspondingly large uncertainty in $\sigma_*$,
as $\sigma$ becomes indendent of the galaxies' luminosity in this
limit.

On the other hand, the log-likelihood function exhibits a pronounced
maximum as a function of $z_m'$, as seen in the lower panel of Fig.\ts
7. The true (input) value is $z_m'=0.1$, and the allowed range of
$z_m'$ is about $0.05\lesssim z_m'\lesssim 0.15$. The corresponding
range in $\sigma_*$ is then $145\lesssim \sigma_*/\hbox{km/s}\lesssim
165$.

We conclude from this that the method for analyzing
galaxy-galaxy lensing introduced here is not very sensitive to
variations of parameters and that, in fact, the likelihood ratio test
can actually constrain the range of model parameters from the data. 

\subs{4.3 Redshift information}
Up to now we have assumed that the apparent magnitude is the only
observable which determines the probability distribution of the
redshift of a galaxy, using the distribution function (2.7). However,
it is conceivable that additional information on the galaxy images can
be obtained, most noticeably the color, which might be used to
constrain the redshift interval. For example, Connolly et al.\ts
(1995) have demonstrated that the redshifts for galaxies
with $B_J\le 22.5$ can be estimated from four-color photometry
to an accuracy of $\Delta z\lesssim 0.05$.

We shall now
demonstrate that $s_*$ and $\sigma_*$ can be determined much more
accurately, if such redshift estimates are available.
Again we use the same data as for Fig.\ts 6, but assume that
the redshift of the galaxies have been estimated with a fractional
accuracy of $\eps$. For simplicity
we assume that the redshift of a galaxy lies within
the interval $(1-\eps)z_{\rm true}\le z\le (1+\eps)z_{\rm true}$,
of the true redshift.  We then use this
redshift interval in the Monte-Carlo integration of $\ave{\gamma_i}$
and $\sigma_{\gamma,i}^2$ which enter the log-likelihood function
(2.20). 
\xfigure{8}{Log-likelihood contours for the same parameters as for the
lower right panel in Fig.\ts 6, i.e., $\theta_{\rm min}=5\arcsec$,
$m_{\rm shape}=23.5$, but with `redshift information' as described in
the text. The left (right) panel corresponds to
$\eps=0.05$($\eps=0.5$). Note that 
the region of the $(s_*,\sigma_*)$-plane displayed here is smaller
than in the previous figures}
{fig8.tps}{14}
In Fig.\ts8 we have plotted the log-likelihood contours for the same
parameters as for the lower right panel of Fig.\ts6, except that we
used the redshift distribution just mentioned, with $\eps=0.05$ (left
panel) and $\eps=0.5$ (right panel). For the same number
of galaxies the contours are now tighter than
those in Fig.\ts6, i.e., $\sigma_*$ and a
lower limit on $s_*$ can be much better determined if redshift
information is included. In fact, in the present case we can actually
also determine an upper limit on $s_*$. The comparison of the two
panels in Fig.\ts 8 shows that even an imprecise
redshift estimate from multi-color photometry considerably
improves the estimates of the lens parameters.

\sec{5 Discussion}
In this paper we have developed an efficient, quantitative analysis
method for galaxy-galaxy lensing, an effect which has been
observationally detected by
BBS. The method is based on a maximum likelihood approach which
accounts for all of the information available from observation.
It accounts not only for the image ellipticities, but uses
the actual relative positions and magnitudes for all surrounding
galaxies which might be potential lenses. From
synthetically generated data sets we have shown that our method can
be used to determine statistical properties of galaxies, such as the
velocity dispersion $\sigma_*$ of an $L_*$-galaxy or its
characteristic tidal cut-off radius. Even for moderately large galaxy
samples, the accuracy of this determination is quite high. For
example, with only 10 sufficiently deep WFPC2 images can $\sigma_*$ be
determined with a statistical accuracy of about 10\%. Alternatively,
high-quality images with a 4-meter class telescopes taken with a
non-too-small field-of-view can yield the required galaxy number in a
few nights. After all, it seems from our analysis that the requirement
on the amount of data is not very demanding, except that the
systematic effects which affect the measured ellipticities on
ground-based images has to be understood sufficiently well.
 
We have also
demonstrated that other statistical properties of the galaxy
distribution can be constrained with this approach, such as the mean
redshift as a function of apparent magnitude. Finally we showed that
an approximate determination of galaxy redshifts increases the
accuracy of our method considerably, so that multi-color photometry of
the galaxies will be very useful.

Several systematic effects may affect the conclusions derived
here. For example, galaxies cannot be realistically described by
spherical potentials, and the effects of an elliptical projected mass
density on the likelihood function should be tested. Furthermore,
since the typical shear caused by any individual foreground galaxy is
at most a few percent, the influence of a large-scale cosmological
shear, or the effect of a cluster not too far from the line-of-sight,
should be studied. Whereas these problems have to be kept in mind (and
we will consider them in a later paper), it is unlikely that these
systematics will significantly modify the results of the current
investigation. In contrast, it may be possible that the detection of a
cosmic shear through image distortions will be easier if the
galaxy-galaxy lensing effects are statistically removed from the
data. 

\noindent
\SFB

\def\ref#1{\vskip1pt\noindent\hangindent=40pt\hangafter=1 {#1}\par}
\sec{References}
\ref{Blandford, R.D. \& Jaroszy\'nski, M. 1981, ApJ 246, 1.}
\ref{Blandford, R. \& Narayan, R. 1986, ApJ 310, 568.}
\ref{Blandford, R.D., Saust, A.B., Brainerd, T.G. \& Villumsen, J.V.\ 1991,
         MNRAS 251, 600.}
\ref{Bonnet, H., Fort, B., Kneib, J.-P., Mellier, Y. \& Soucail, G. \
         1993, A\&A 280, L7.}
\ref{Bonnet, H., Mellier, Y. \& Fort, B.\ 1994, ApJ 427, L83.}
\ref{Brainerd, T.G., Blandford, R.D. \& Smail, I.\ 1995, ApJ, in press
(BBS).}
\ref{Connolly, A.J., Csabai, I., Szalay, A.S., Koo, D.C., Kron,
         R.G. \& Munn, J.A.\ 1995, preprint.}
\ref{Fahlman, G., Kaiser, N., Squires, G. \& Woods, D.\ 1994, ApJ 437, 56.}
\ref{Fort, B. \& Mellier, Y.\ 1994, A\&AR 5, 239.}
\ref{Fort, B., Mellier, Y., Dantel-Fort, M., Bonnet, H. \& Kneib, J.-P.\
         1995, A\&A, in press.}
\ref{Kaiser, N.\ 1992, ApJ 388, 272.}
\ref{Kaiser, N. \& Squires, G.\ 1993, ApJ 404, 441.}
\ref{Kovner, I. \& Milgrom, M. 1987, ApJ 321, L113.}
\ref{Kristian, J.\ 1967, ApJ 147, 864.}
\ref{Miralda-Escud\'e, J.\ 1991, ApJ 370, 1.}
\ref{Mould, J., Blandford, R., Villumsen, J., Brainerd, T., Smail, I.,
         Small, T. \& Kells, W.\ 1994, MNRAS 271, 31.}
\ref{Schneider, P.\ 1995, A\&A 302, 639.}
\ref{Schneider, P., Ehlers, J. \& Falco, E.E.\ 1992, {\it Gravitational
         lenses}, Springer: New York (SEF).}
\ref{Schneider, P. \& Seitz, C.\ 1995, A\&A 294, 411.}
\ref{Seitz, C., Kneib, J.-P., Schneider, P. \& Seitz, S.\ 1995, A\&A
            (submitted).}
\ref{Smail, I., Hogg, D., Yian, L. and  Cohen, J.\ 1995, ApJ, 449, L1}
\ref{Smail, I., Ellis, R.S., Fitchett, M.J. \& Edge, A.C.\ 1995, MNRAS
273, 277.}
\ref{Squires, G.\ et al.\ 1995, preprint.}
\ref{Tyson, J.A., Valdes, F., Jarvis, J.F. \& Mills Jr., A.P. 1984, ApJ
         281, L59.}
\ref{Tyson, J.A., Valdes, F. \& Wenk, R.A. 1990, ApJ 349, L1.}
\ref{Villumsen, J.V.\ 1995b, MNRAS, submitted.}
\ref{Webster, R.L. 1985, MNRAS 213, 871.}
\ref{Zaritzky, D. \& White, S.~D.~M.\ 1994, ApJ, 435, 599}

\vfill\eject\end